\documentclass{iau}

\usepackage{amsmath}
\usepackage{graphicx}
\usepackage{multirow}
\usepackage{pgf,tikz}
\usepackage{caption}
\definecolor{bleufonce}{rgb}{0.35, 0.0, 0.55} 

\definecolor{carotte}{rgb}{0.96, 0.4,0.11}

\begin{document}

\lefttitle{Rosu et al.}
\righttitle{Apsidal motion in O-star binaries: GENEC rotating binary models put to the $k_2$-test}

\jnlPage{1}{7}
\jnlDoiYr{2025}
\doival{}

\aopheadtitle{Proceedings IAU Symposium}
\editors{A. Wofford,  N. St-Louis, M. Garcia \&  S. Simón-Díaz, eds.}

\title{Apsidal Motion in O-Star Binaries: \\GENEC rotating binary models put to the $\mathbf{k_2}$-test}

\author{S.~Rosu$^1$, L.~Sciarini$^1$, S.~Ekström$^1$, \& R.~Hirschi$^{2,3}$}
\affiliation{$^1$Département d’Astronomie, Université de Genève, Chemin Pegasi 51, CH-1290 Versoix, Switzerland.\\
$^2$Astrophysics Research Centre, Lennard-Jones Laboratories, Keele University, Keele ST5 5BG, UK.\\
$^3$Kavli IPMU (WPI), The University of Tokyo, 5-1-5 Kashiwanoha, Kashiwa 277-8583, Japan}

\begin{abstract}
Unveiling massive stars’ internal structure and the physical origin and efficiency of the internal mixing processes? It is now possible using the apsidal motion rate in close eccentric binaries! The apsidal motion rate depends on the tidal interactions occurring between the stars and is proportional to $k_2$, a measure of the star’s inner density profile. Confronting standard stellar models with observations reveals the famous $k_2$-discrepancy: models predict too high a $k_2$ for the stars, that is to say, stars with too low a density contrast between their core and envelope. We built bespoke GENEC stellar evolution models including tidal mixing for the twin massive binary HD\,152248. The models reveal the instabilities allowing to reproduce the stellar density profiles: advecto-diffusive models better reproduce $k_2$ than magnetic models. A large overshooting is necessary to converge towards the observed $k_2$, yet alone is not sufficient. While a change in metallicity or mass-loss rate has no significant impact on $k_2$, a larger initial helium abundance allows us to better reproduce the $k_2$. Yet, a super-solar helium abundance is not observationally supported. Our analyses highlight the need for a process in the stars that slows down the radial expansion.
\end{abstract}


\maketitle

\section{The beauty of $\mathbf{k_2}$}
\noindent In close eccentric binaries tidal interactions lead to dynamical deformations of the stellar surfaces as the stars orbit each other. The ensuing non-spherical stellar gravitational fields in turn induce secular changes of the orbital parameters. The most prominent manifestation of this effect is the slow precession of the orbit major axis, known as apsidal motion. The rate of this motion $\dot\omega$ is directly related to the internal structure constants of each star, $k_2$, a measure of the mass distribution between the star’s core and external layers \citep{shakura85}, mainly affected by the convective core size and internal mixing efficiency. Measuring $\dot\omega$ hence provides a powerful and independent diagnosis of the otherwise difficult to constrain density stratification inside stars and offers a unique means to reveal stars’ structure and evolution \citep{claret10}. This challenging technique has been known for decades, but was data and models starved until recently. Now that accurate spectroscopic and photometric data can be obtained, and that the evolution of binaries can be modelled with detailed binary stellar evolution codes, the full potential of apsidal motion can finally be exploited.
%
The rate of apsidal motion $\dot\omega$ is made of a Newtonian contribution $\dot\omega_\text{N}$ -- usually the dominant one -- that accounts for tidal deformations and stellar rotation, and a general relativistic correction $\dot\omega_\text{GR}$:
{\small{
\begin{equation}
\label{eqn:AM}
\begin{aligned}
\dot\omega_\mathrm{N} = &\frac{2\pi}{P_\text{orb}} \Bigg[15f(e)\left\{k_{2,1}q \left(\frac{R_1}{a}\right)^5 + \frac{k_{2,2}}{q} \left(\frac{R_2}{a}\right)^5\right\} \\
& + g(e) \Bigg\{ k_{2,1} (1+q) \left(\frac{R_1}{a}\right)^5 \left(\frac{P_\mathrm{orb}}{P_\text{rot,1}}\right)^2+ k_{2,2}\, \frac{1+q}{q} \left(\frac{R_2}{a}\right)^5 \left(\frac{P_\mathrm{orb}}{P_\text{rot,2}}\right)^2 \Bigg\}  \Bigg],\\
\end{aligned}
\end{equation}
\begin{equation}
\dot\omega_\mathrm{GR}  = \left(\frac{2\pi}{P_\mathrm{orb}}\right)^{5/3}\frac{3(G(m_1+m_2))^{2/3}}{c^2 (1-e^2)},
\end{equation}}}
\noindent where  $q=M_2/M_1$ is the mass ratio, $P_\text{orb}$ the orbital period, $a$ the semi-major axis of the orbit, $R_\star$ the radius, $k_{2,\star}$ the internal structure constant, and $P_{\mathrm{rot,}\star}$ the rotational period of the considered star, $f(e)$ and $g(e)$ functions of the eccentricity of the orbit which expressions are given in e.g., \citet{rosu21}, $G$ the gravitational constant, and $c$ the speed of light. $k_2$ is an algebraic function of $\eta_2$, solution of the Clairaut-Radau differential equation that depends on the stellar density profile \citep{hejlesen87}. $k_2$ is a monotonic decreasing function with time \citep[their Figs 8\&9]{rosu21}; its value is mostly impacted by the density contrast at the border between the convective core and radiative envelope of the star \citep[their Fig.\,1]{rosu20a}.

\section{The $\mathbf{k_2}$ discrepancy\label{sect:k2dis}}
\noindent The $k_2$ discrepancy is the systematic difference between the $k_2$ values obtained from stellar evolution models and observations for massive stars: the former is systematically larger than the latter \citep{rosu20a, rosu22a, rosu22b}. It means that models predict stars with too low a density contrast, that is to say, observations suggest that the core is more contracted and the envelope more extended.
In this paper, we investigate this $k_2$ discrepancy in the benchmark twin binary HD\,152248. Its twin nature -- meaning that the two stars share exactly the same stellar properties -- allows us to directly get $k_2$ from the observations, assuming $k_2 \equiv k_{2,1} = k_{2,2}$ in Eq.\,\eqref{eqn:AM}. The observational properties of HD\,152248 are given in Table\,\ref{table:HD152248} \citep{rosu20b}. Among these properties, the radius is particularly important: it appears to the fifth power in Eq.\,\eqref{eqn:AM}, so it is important that the models do reproduce it within its error bars.
The metallicity of the NGC\,6231 cluster to which HD\,152248 belongs is slightly super-solar: $Z=0.018\pm0.003$ \citep{dias21}. The initial He abundance of the cluster is poorly-constrained as different studies achieve conflicting results: \citet{lennon83} derived a slightly super-solar while \citet{mathys02} a slightly sub-solar abundance.

\begin{table}
\caption{Stellar and orbital parameters of HD\,152248 \citep{rosu20b}.\label{table:HD152248}}
\centering
\begin{tabular}{l l l l l}
\hline
Parameter & Symbol & Value \\
\hline
Mass & $M_\star$ ($M_\odot$) & $29.5\pm0.5$\\
Radius & $R_\star$ ($R_\odot$) & $15.07\pm 0.12$\\
Effective temperature & $T_{\text{eff},\star}$ (K) & $34000\pm1000$\\
Bolometric luminosity & $L_{\text{bol},\star}$ ($L_\odot$) & $2.73\pm0.32 \times^5$\\
Internal stellar structure constant & $k_{2,\star}$ & $0.0010 \pm 0.0001$\\
Surface velocity & $v_{\text{surf},\star}$ (km\,s$^{-1}$) & $148\pm13$\\
Mass-loss rate & $\dot{M}$ ($M_\odot$\,yr$^{-1}$) & $\le 8\times 10^{-7}$ \\
Nitrogen surface abundance & N/H (nb) & $0.6-2.2 \times 10^{-4}$ \\
Orbital period & $P_\text{orb}$ & $5.81650\pm0.00002$\\
Orbital eccentricity & $e$ & $0.130\pm0.002$\\
Apsidal motion rate & $\dot\omega$ ($^\circ$\,yr$^{-1}$) & $1.843\pm0.083$\\
\hline
\end{tabular}
\end{table}

\section{GENEC stellar evolution models}
\noindent We built stellar models with the Geneva code \citep[GENEC,][]{eggenberger08}, adopting two rotational mixing implementations, the hydrodynamic and magnetic schemes. We compare single and binary star models computed with the same internal physics. For the binary models, dynamical tides and equilibrium tides acting on small subsurface convective zones are included as described in \citet{song13, sciarini24, sciarini25}. 
We adopted the mass-loss rate of \citet{krticka24} as it best reproduces observed stars \citep{brands22}. We adopted the metallicity of the cluster and a solar initial He abundance. For the different models, we fine-tuned the initial mass $M_\text{init}$ such that $M(R_\star) = M_\star$\footnote{As outlined in Sect.\,\ref{sect:k2dis}, if we get the radius wrong, everything else is wrong!}. For the single-star models, we adopted the same approach for the surface velocity: we fine-tuned the initial velocity $v_\text{ini}$ such that $v_\text{surf} (R_\star) = v_{\text{surf},\star}$. For binary models, we initialised the system synchronised to the orbital motion.
We tested values of the overshooting parameter $\alpha_\text{ov}$ from 0.5 to 1.0. These high values are motivated by several independent studies showing the need for larger convective boundary mixing in massive stars than anticipated and predicted from lower-mass stars in order to: reproduce the main-sequence width in the Hertzsprung-Russell (HR) diagram of stellar clusters at high masses \citep{castro14, baraffe23}, explain the properties of massive eclipsing binaries with 1D stellar models \citep{tkachenko20, higgins19}, 2D and 3D hydrodynamic simulations \citep{baraffe23,mao24}, and 3D non-perturbative modelling \citep{fellay24}, justify the convective core entrainment in 321D-guided 1D stellar models \citep{scott21}, and explain asteroseismic gravity modes of single stars \citep{aerts21}.

\subsection{Single star models}
\begin{figure}
\centering
\includegraphics[width=0.49\linewidth]{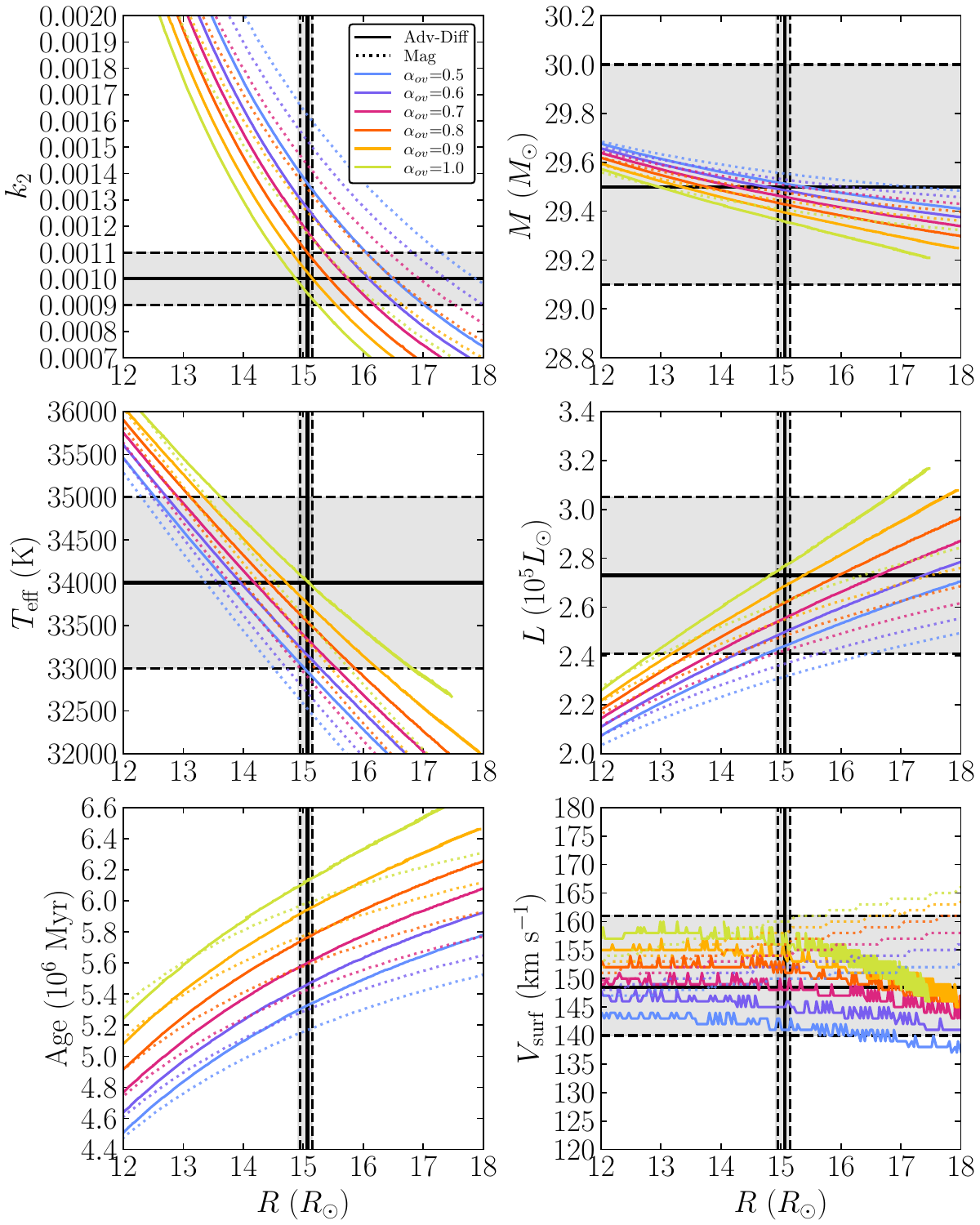}
\includegraphics[width=0.49\linewidth]{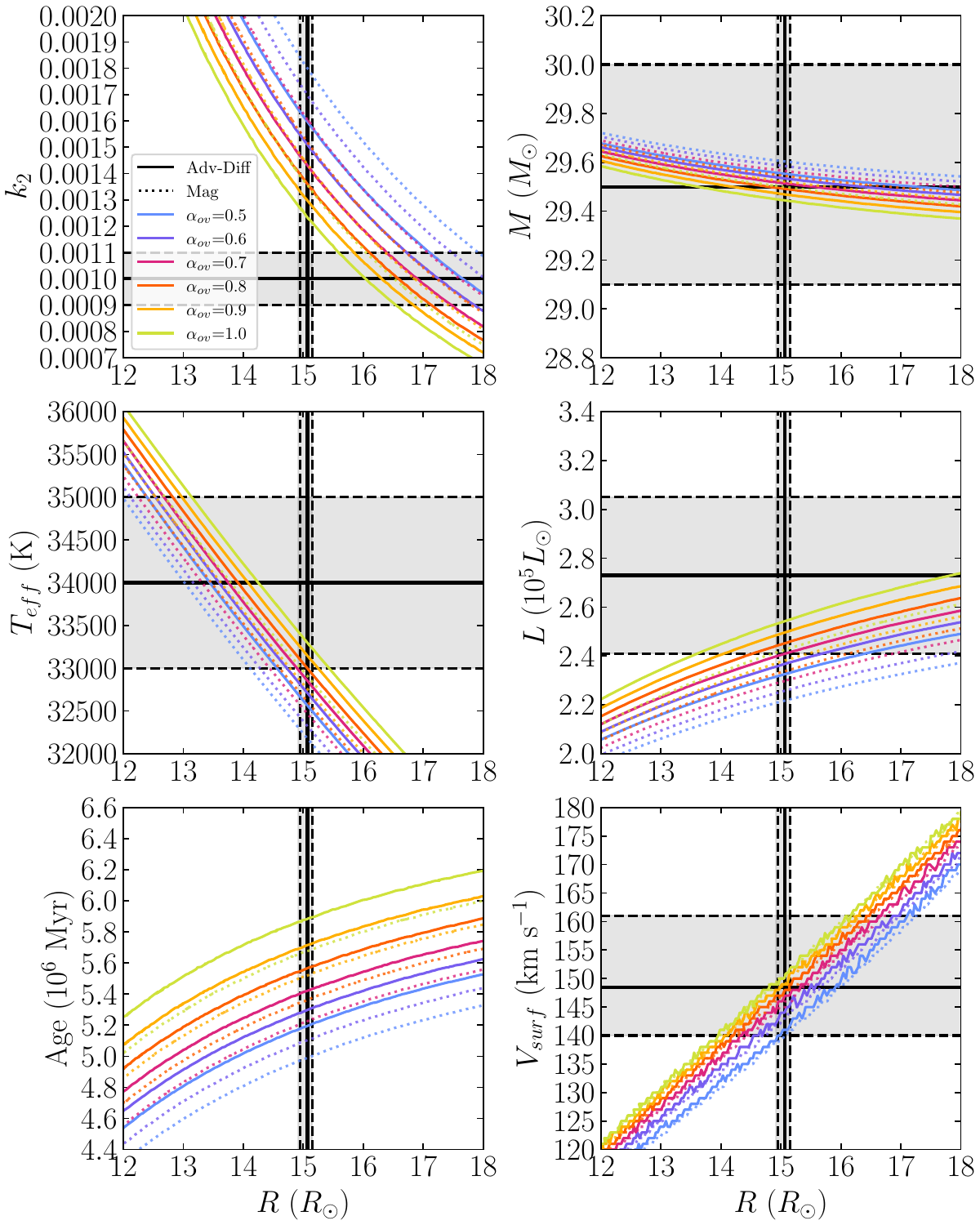}
\caption{Evolution of stellar parameters with radius for single (columns 1\&2) and binary (columns 3\&4) star models, with observational values and error bars (plain and dashed lines).}
\label{fig:single}
\end{figure}

\noindent Single star models with $M_\text{init} = 30.2\,M_\odot$ are shown in Fig.\,\ref{fig:single} (columns 1\&2). The $k_2$ discrepancy is clearly observed: Only advecto-diffusive models with $\alpha_\text{ov} \ge 0.8$ reproduce all parameters ($M, R, T_\text{eff}, L, k_2$). The need for large $\alpha_\text{ov}$ is explained as follows: A larger $\alpha_\text{ov}$ means the star has a larger convective core hence more fuel for burning. Its main-sequence lifetime is  extended and the model is older when it reaches $R_\star$. The star has had more time to contract its interior and extend its envelope and has therefore a smaller $k_2$ when reaching $R_\star$. 

\subsection{Binary star models}
\begin{figure}
\centering
\includegraphics[width=0.75\linewidth]{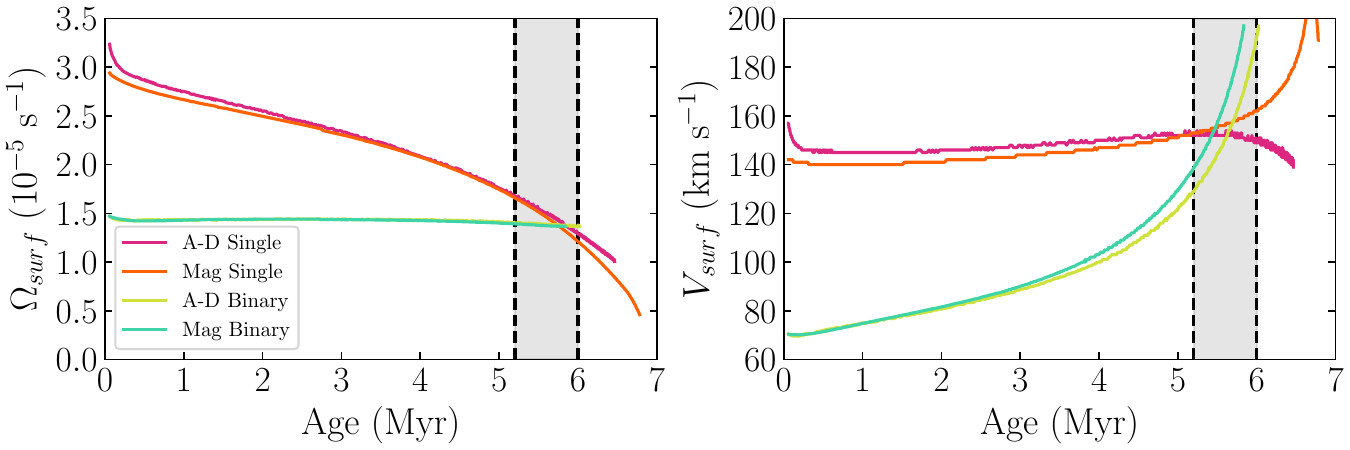}
\caption{Evolution of $\Omega_\text{surf}$ and $v_\text{surf}$ with time for single and binary models.}
\label{fig:vsurf}
\end{figure}

\noindent Binary star models are shown in Fig.\,\ref{fig:single} (columns 3\&4). Not only the models do not reproduce the $k_2$ at $R_\star$ -- even with large $\alpha_\text{ov}$ -- but also they are worse at reproducing the position in the HR diagram. If the single star models are apparently better at reproducing the current properties of the stars, it is because we chose $v_\text{ini}$ such that $v_\text{surf} (R_\star) = v_{\text{surf},\star}$. During its evolution, a single star model sees its surface angular velocity $\Omega_\text{surf}$ decrease and its radius increase in time; Its $v_\text{surf}$ is thus constant in time (Fig.\,\ref{fig:vsurf}). In reality, the star has a companion. Its $\Omega_\text{surf}$ is fixed by the orbital motion (the binary is pseudo-synchronised) while its radius is increasing in time. Its $v_\text{surf}$ thus increases in time (Fig.\,\ref{fig:vsurf}). Compared to a single star model, a binary star model has lower $\Omega_\text{surf}$ and $v_\text{surf}$ throughout its lifetime until reaching $R_\star$, thus has been less mixed during its evolution. Consequently, a binary model has a larger $k_2$ when reaching $R_\star$ than its single-star counterpart. The conclusion is not that the single star models are better are reproducing the stellar properties, but that it is important to take into account the binary interactions by computing binary models. Yet, something is missing in the binary models to reproduce the stellar properties. We investigate several possibilities hereafter. 

\begin{figure}
\begin{minipage}[b]{0.49\linewidth}
\centering
\includegraphics[width=\linewidth]{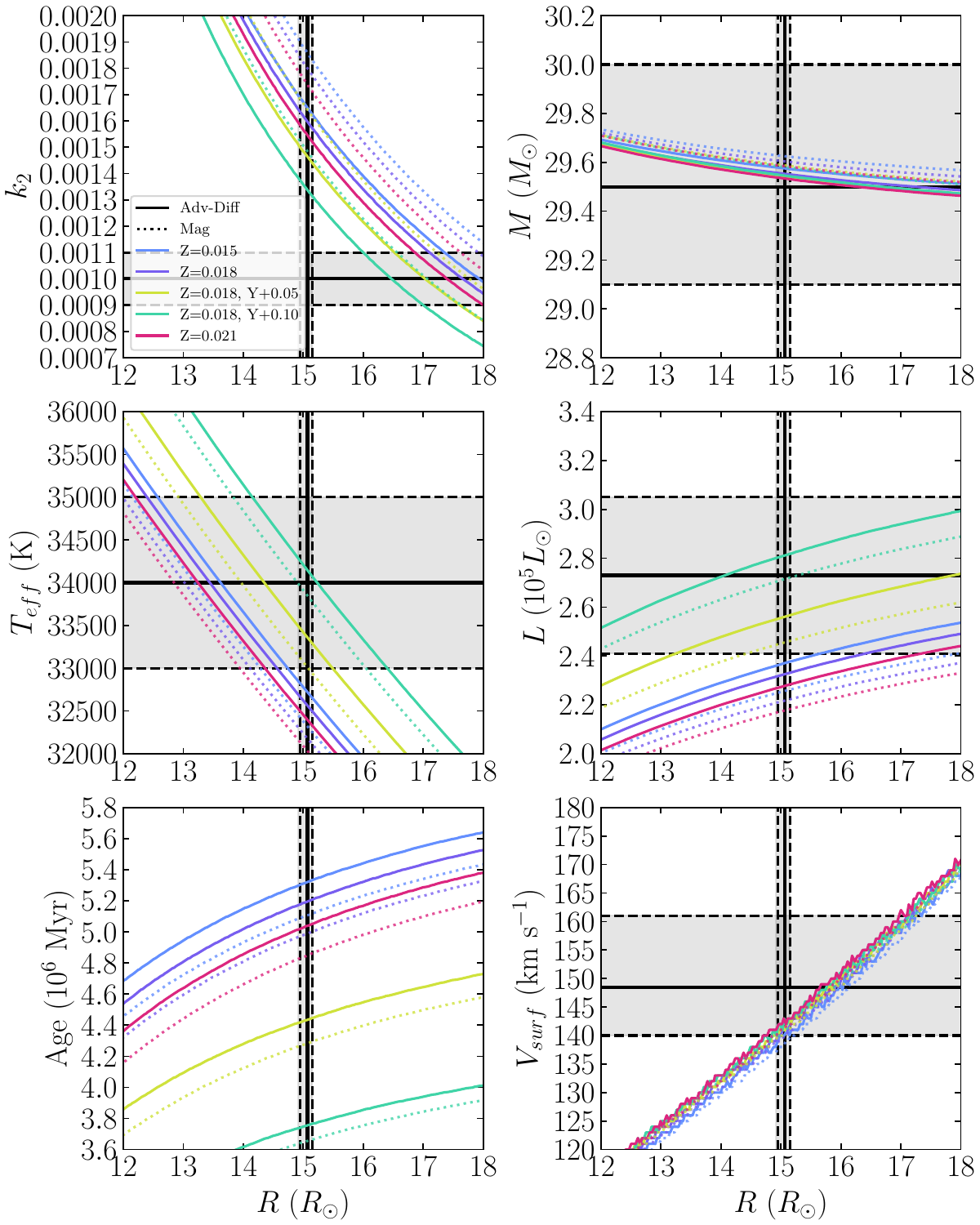}
\caption{Evolution of stellar parameters with $R$ for binary models with different $Z$ and $He$.}
\label{fig:ZHe}
\end{minipage}
\hfill
\begin{minipage}[b]{0.49\linewidth}
\centering
\includegraphics[width=\linewidth]{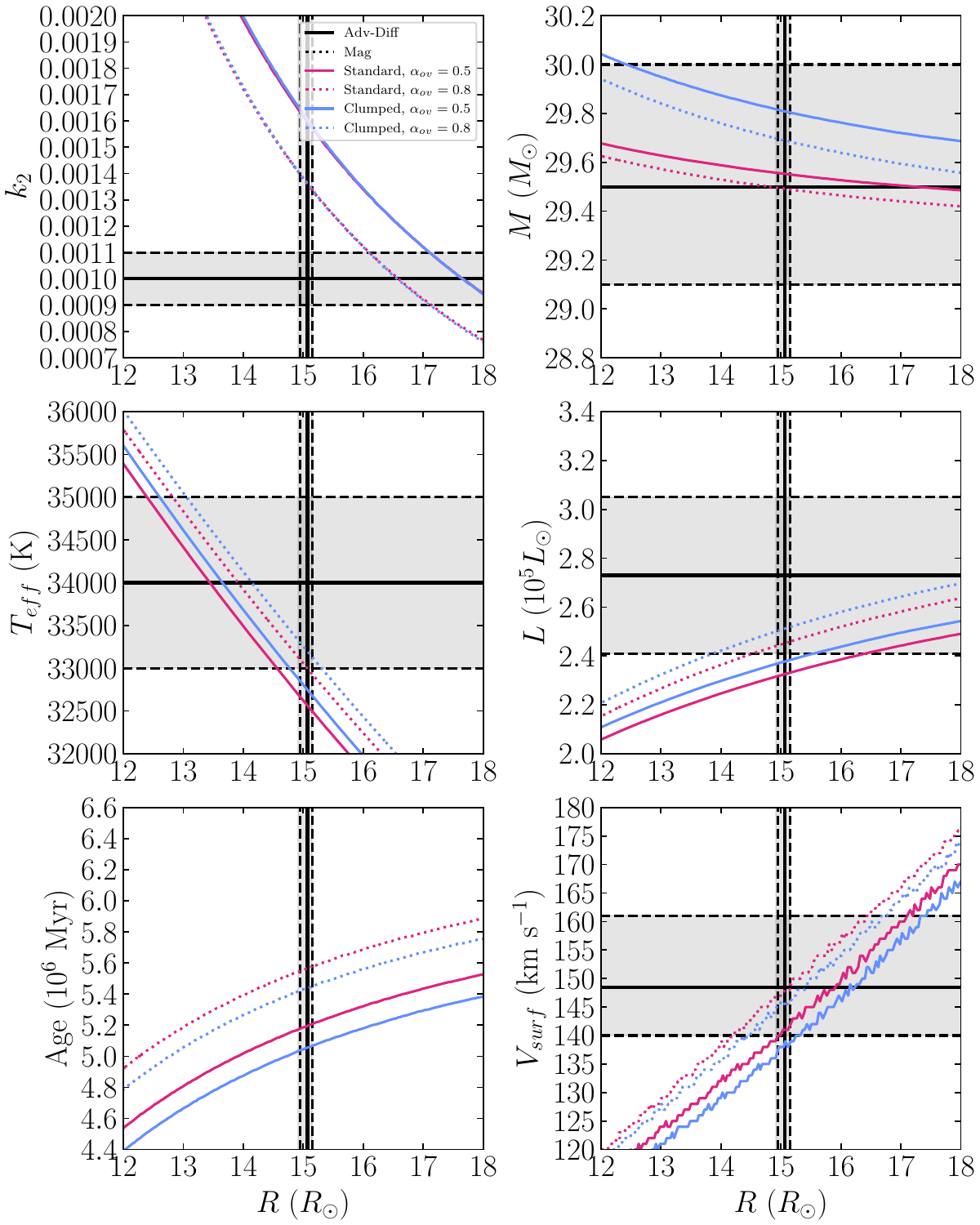}
\caption{Evolution of stellar parameters with $R$ for binary models with larger mass-loss rate.}
\label{fig:mdot}
\end{minipage}
\end{figure}

\noindent \textbf{Could it be the metallicity or initial helium abundance?}
We tested the influence of the metallicity and the initial He abundance on the models. We varied $Z$ within its error bars: an increase in $Z$ leads to a decrease in $k_2$ but also a decrease in $T_\text{eff}$ and $L$ (Fig.\,\ref{fig:ZHe}). Yet, the impact is very limited. An increase in He leads to a decrease in $k_2$ and an increase in $T_\text{eff}$ and $L$ (Fig.\,\ref{fig:ZHe}). While it solves partially the problem, it is not supported observationally.

\noindent \textbf{Could it be the mass-loss rate?}
Winds are clumped, and if clumping starts close to the photosphere, theoretical mass-loss rates are underestimated by a factor $f^{0.25}$, with $f = 8-10$ \citep{krticka24}. We computed stellar models with \citet{krticka24} mass-loss rate multiplied by 1.8\footnote{It should be an upper limit (Krticka, priv. com.).}. The models have a slightly larger $M_\text{ini} = 32.0 M_\odot$, thus slightly larger $T_\text{eff}$ and $L$ at $R_\star$ (Fig.\,\ref{fig:mdot}). Yet, $k_2$ does not change compared to non-clumped models and the $k_2$-discrepancy is still not solved. The reason is that the star is initially slightly more massive so its $k_2$ decreases faster. However, it reaches $R_\star$ earlier thus has not had time to evolve as much as the non-clumped models. Its density contrast is not as pronounced, which compensates the faster decrease in $k_2$. In conclusion, even if mass-loss rate is underestimated, it has no significant impact on the models.



\section{GENEC models put to the $\mathbf{k_2}$-test: Status quo}
\noindent With the best up-to-date physics included, GENEC models do show the $k_2$ discrepancy for HD\,152248. The standard binary models predict too low $T_\text{eff}$ and $L$ but too high a $k_2$. Only an increase in $\alpha_\text{ov}$ allows the models to get closer to reproduce the observations. A change in the metallicity has a non-significant impact on the models. An increase in the initial He abundance allows to better reproduce the observations; It is however not observationally supported. Clumped binary models with larger mass-loss rates slightly improve on $T_\text{eff}$ and $L$ predictions -- mostly because $M_\text{ini}$ is larger. Yet, the impact on $k_2$ is non-significant. Therefore, even if the mass-loss rate of our standard models is underestimated, it has no impact on our conclusions that models predict too high a $k_2$. 
In summary, our investigations suggest that we need a mechanism that slows down the stellar radius increase in time. In the future, we will increase by an order of magnitude the number of binaries studied this way in a wide range of stellar and orbital parameters.



\begin{thebibliography}{}
\bibitem[Aerts(2021)]{aerts21}
Aerts, C. 2021, RvMP, 93, 1
\bibitem[Baraffe et al.(2023)]{baraffe23}
Baraffe, I., Clarke, J., Morison, A., et al. 2023, MNRAS, 519, 5333
\bibitem[Brands et al.(2022)]{brands22}
Brands, S. A., de Koter, A., Bestenlehner, J. M., et al. 2022, A\&A, 663, A36
\bibitem[Castro et al.(2014)]{castro14}
Castro, N., Fossati, L., Langer, N., et al. 2014, A\&A, 570, L13
\bibitem[Claret \& Giménez(2010)]{claret10}
Claret, A., \& Giménez, A. 2010, A\&A, 519, A57
\bibitem[Dias et al.(2021)]{dias21}
Dias, W. S., Monteiro, H., Moitinho, A., et al. 2021, MNRAS, 504, 356
\bibitem[Eggenberger et al.(2008)]{eggenberger08}
Eggenberger, P., Meynet, G., Maeder, A., et al. 2008, Ap\&SS, 316, 43
\bibitem[Fellay et al.(2024)]{fellay24}
Fellay, L., Dupret, M.-A., \& Rosu, S. 2024, A\&A, 683, A210
\bibitem[Hejlesen(1987)]{hejlesen87}
Hejlesen, P. M. 1987, A\&AS, 69(2), 251
\bibitem[Higgins \& Vink(2019)]{higgins19}
Higgins, E. R., \& Vink, J. S. 2019, A\&A, 622, A50
\bibitem[Krtička et al.(2024)]{krticka24}
Krtička, J., Kubát, J., \& Krtičková, I. 2024, A\&A, 681, A29
\bibitem[Lennon \& Dufton(1983)]{lennon83}
Lennon, D. J., \& Dufton, P. L. 1983, MNRAS, 203, 443
\bibitem[Mao et al.(2024)]{mao24}
Mao, H., Woodward, P., Herwig, F., et al. 2024, ApJ, 975, 271
\bibitem[Mathys et al.(2002)]{mathys02}
Mathys, G., Andrievsky, S. M., Barbuy, B. et al. 2002, A\&A, 387, 890
\bibitem[Rosu(2021)]{rosu21}
Rosu, S. 2021, BSRSL, 90, 1
\bibitem[Rosu et al.(2020a)]{rosu20a}
Rosu, S., Noels, A., Dupret, M.-A., et al. 2020a, A\&A, 642, A221
\bibitem[Rosu et al.(2020b)]{rosu20b}
Rosu, S., Rauw, G., Conroy, K. E., et al. 2020b, A\&A, 635, A145
\bibitem[Rosu et al.(2022a)]{rosu22a}
Rosu, S., Rauw, G., Farnir, M., et al. 2022a, A\&A, 660, A120
\bibitem[Rosu et al.(2022b)]{rosu22b}
Rosu, S., Rauw, G., Nazé, Y., et al. 2022b, A\&A, 664, A98
\bibitem[Sciarini et al. (2024)]{sciarini24}
Sciarini, L., Ekström, S., \& Eggenberger, P. 2024, A\&A, 681, L1
\bibitem[Sciarini et al.(2025)]{sciarini25}
Sciarini, L., Rosu, S., Ekström, S., et al. 2025, A\&A, submitted
\bibitem[Scott et al.(2021)]{scott21}
Scott, L. J. A., Hirschi, R., Georgy, C., et al. 2021, MNRAS, 503, 4208
\bibitem[Shakura(1985)]{shakura85}
Shakura, N. I. 1985, Sov. Astron. Lett., 11, 224
\bibitem[Song et al.(2013)]{song13}
Song, H. F., Maeder, A., Meynet, G., et al. 2013, A\&A, 556, A100
\bibitem[Tkachenko et al.(2020)]{tkachenko20}
Tkachenko, A., Pavlovski, K., Johnston, C., et al. 2020, A\&A, 637, A60
\end{thebibliography}
\end{document}